\input harvmac
\input epsf

\noblackbox 

\Title
 {\vbox{
 \baselineskip12pt
 \hbox{HUTP-01/A057}
 \hbox{TIFR/TH/01-45}
 \hbox{hep-th/0111155}\hbox{}\hbox{}
}}
{\vbox{
 \centerline{$tt^*$ Geometry
}\vglue .75cm
  \centerline{and }
 \vglue .75cm
 \centerline{ Closed String Tachyon Potential}
 }}
 \medskip
\centerline{Atish Dabholkar$^{1,2}$ and Cumrun Vafa$^{2}$}
\vskip 1 cm

\centerline{\ $^1$ Department of Theoretical Physics, Tata
Institute of Fundamental Research, Mumbai 400005, India}

\centerline{\ $^2$ Jefferson Physical Laboratory, Harvard University,
Cambridge, MA 02138, USA}
\medskip
\bigskip

\vskip .1in\centerline{\bf Abstract}

We propose a  closed string tachyon action including kinetic
and potential terms for
non-supersymmetric orbifolds.  The action
is given in terms of solutions to $tt^*$ equations which
captures the geometry of vacua of the 
corresponding $N=2$ worldsheet theory.   In certain cases
the solutions are well studied.  In case of tachyons of
${\bf C}/Z_n$,  solutions to affine
toda equations determine the action. We study the particular case
of ${\bf C}/Z_3\rightarrow {\bf C}$ in detail and find that
 the Tachyon action is determined in terms of
a solution to Painleve III equation.

 \smallskip
\Date{November 2001}

\lref\beret{M. Bershadsky, A Johansen, T. Pantev, V. Sadov and C. Vafa,
`` F-theory, Geometric Engineering and N=1 Dualities,''
Nucl.Phys.{\bf B505} (1997)  153-164 [hep-th/9612052]; C. Vafa and
B. Zwiebach, ``N=1 Dualities of SO and USp Gauge Theories
and T-Duality of String Theory,'' Nucl.Phys. {\bf B506}
(1997) 143-156 [hep-th/9701015].}
\lref\ov{H. Ooguri and C. Vafa,``Geometry of N=1 Dualities in Four
Dimensions,'' Nucl.Phys. {\bf B500} (1997) 62-74 [hep-th/9702180].}

\def\fund{  \> {\vcenter  {\vbox
              {\hrule height.6pt
               \hbox {\vrule width.6pt  height5pt
                      \kern5pt
                      \vrule width.6pt  height5pt }
               \hrule height.6pt}
                         }
                   } \>
           }

\batchmode
  \font\bbbfont=msbm10
\errorstopmode
\newif\ifamsf\amsftrue
\ifx\bbbfont\nullfont
  \amsffalse
\fi
\ifamsf
\def\IR{\hbox{\bbbfont R}}

\def\IZ{\hbox{\bbbfont Z}}
\def\IF{\hbox{\bbbfont F}}
\def\IP{\hbox{\bbbfont P}}
\else
\def\IR{\relax{\rm I\kern-.18em R}}
\def\IZ{\relax\ifmmode\hbox{Z\kern-.4em Z}\else{Z\kern-.4em Z}\fi}
\def\IF{\relax{\rm I\kern-.18em F}}
\def\IP{\relax{\rm I\kern-.18em P}}
\fi

\lref\hiv{K. Hori, A. Iqbal and C. Vafa, ``D-Branes And Mirror
Symmetry,'' [hep-th/0005247].}

\lref\klebwi{I. Klebanov and E. Witten, ``Superconformal Field Theory on
Threebranes at a Calabi-Yau Singularity,''
Nucl.Phys. {\bf B536} (1998) 199-218 [hep-th/9807080].}

\lref\lnv{A. Lawrence, N. Nekrasov and C. Vafa, ``On Conformal Theories in
Four Dimensions,'' Nucl.Phys. {\bf B533} (1998) 199-209 [hep-th/9803015].}

\lref\su{T.~Suyama,
``Properties of string theory on Kaluza-Klein Melvin background,''
arXiv:hep-th/0110077\semi ``Melvin background in heterotic theories,''
arXiv:hep-th/0107116\semi ``Closed string tachyons in 
non-supersymmetric heterotic theories,''
JHEP {\bf 0108}, 037 (2001)
[arXiv:hep-th/0106079].}

\lref\mn{J. Maldacena and C. Nunez,``Towards the large N limit of pure N=1
super Yang Mills,'' Phys.Rev.Lett. {\bf 86} (2001) 588-591 [hep-th/0008001].}

\lref\witli{E. Witten, ``Phases of $N=2$ Theories In Two Dimensions,''
 Nucl.Phys. {\bf B403} (1993) 159-222 [hep-th/9301042].}

\lref\horv{K. Hori and C. Vafa, ``Mirror Symmetry,'' [hep-th/0002222].}

\lref\cfiv{S. Cecotti, P. Fendley, K. Intriligator and C. Vafa, ``A New
Supersymmetric Index,'' Nucl.Phys. B386 (1992) 405-452 [hep-th/9204102].}

\lref\cecov{S. Cecotti and C. Vafa, ``On Classification of N=2
Supersymmetric Theories,'' Commun.Math.Phys. {\bf 158} (1993)
569-644 [hep-th/9211097].}

\lref\harg{J. David, M. Gutperle, M. Headrick and S. Minwalla,
``Closed String Tachyon Condensation on Twisted Circles,''
to appear.}

\lref\dm{M. Douglas and G. Moore, ``D-branes, Quivers, and ALE
Instantons,'' [hep-th/9603167].}

\lref\ks{S. Kachru, E. Silverstein,``4d Conformal Field Theories and
Strings on Orbifolds,'' Phys.Rev.Lett. {\bf 80} (1998) 4855-4858
[hep-th/9802183].}


\newsec{Introduction}

\lref\CostaNW{
M.~S.~Costa and M.~Gutperle,
JHEP {\bf 0103}, 027 (2001)
[arXiv:hep-th/0012072].
}

\lref\fluxb{M. Gutperle and A. Strominger,
``Fluxbranes in String Theory,'' JHEP {\bf 0106}, 035 (2001)
[hep-th/0104136].}

\lref\posi{A. Adams,
J. Polchinski and E. Silverstein, ``Don't Panic! Closed String
Tachyons in ALE Spacetimes,'' hep-th/0108075.}

\lref\vata{C. Vafa, ``Mirror Symmetry and Closed String Tachyon
Condensation,'', hep-th/0111051,}

 \lref\atd{A.
Dabholkar, ``Strings on a Cone and Black
Hole Entropy,'' Nucl. Phys. {\bf B 439}, 650,
1995 [hep-th/9408098]\semi
``Quantum Corrections to Black Hole Entropy in String
Theory,'' Phys. Lett. {\bf B 347} (1995) 222 [hep-th/9409158]\semi 
 ``Tachyon Condensation and Black Hole Entropy,''
[hep-th/0111004].}

Recently the dynamics of closed string tachyons has been addressed
in a number of situations involving non-supersymmetric orbifolds
\refs{\CostaNW ,\fluxb ,\su, \posi , \atd ,\vata
,\harg}.
The aim of this note is to propose a tachyon action for these
cases.  Our proposal
utilizes the $N=2$ supersymmetry
that the worldsheet continues to enjoy upon tachyon condensation
\vata . In particular we use the ground state metric
and a notation of an
 algebraic ``c-function'' defined
in  \ref\cecva{S. Cecotti and C. Vafa,
``Topological Anti-Topological Fusion,'' Nucl. Phys. {\bf  B367} (1991)
359.}\ for $N=2$ theories in 2 dimensions, to construct
the kinetic and potential terms for the Tachyon action.  The geometry
underlying these structures is known as the $tt^*$ geometry.
The ``c-function'' can also be
interpreted as a kind of supersymmetric index $Tr (-1)^F Fe^{-\beta H}$
\ref\ansu{S. Cecotti, P. Fendley, K. Intriligator and C. Vafa,
``A New Supersymmetric Index,'' Nucl. Phys. {\bf B386} (1992) 405.}.

The plan of this note is as follows:  In section 2 we propose
the tachyon action for general $N=2$ worldsheet theories, focusing for
simplicity
on the $C/Z_n$ case.
We 
use the results in \refs{\posi,\atd ,\vata, \cecva }\ to motivate
our proposal.  In section 3
we consider some examples.  The case of ${\bf C/Z_3}\rightarrow
{\bf C}$ is the simplest example and is worked out in great detail.

\newsec{Tachyon Action and $N=2$ Worldsheet Supersymmetry}

In this section we make a proposal for tachyon action involving
both the kinetic and potential terms.  The action
can be stated in full generality
 in terms of purely $N=2$ worldsheet data.   However
for the sake of definiteness
 we will concentrate on the case of ${\bf
C/Z_n}$.  We will see that the natural candidate for the tachyon
potential is given in terms of the highest axial charge of
the Ramond ground states.  The natural candidate for the tachyon
kinetic term is given in terms of the ground state metric of
the Ramond states.  Both of these are captured by 
$tt^*$ equations which we will briefly review.

Tachyon condensation in the context of ${\bf C/Z_n}$ was studied
in \posi .  In order to avoid tachyon in the bulk, one has to
consider odd $n$ and take the $Z_n$ action to be given by
$$g={\rm exp}(4\pi i J/n)$$
where $J$ is the generator of rotation on the complex plane $x$:
$${\rm exp (i\theta J)}\cdot x=e^{i\theta }x.$$
It was argued in \posi\ that giving vev to the
tachyon field will give  a flow to ${\bf C/Z_k}$ theories,
where $k$ is an odd integer satisfying $k<n$ .  In \vata\
it was shown that the worldsheet theory is mirror to
the $N=2$ Landau-Ginzburg theory with a superpotential given by
$$W=u^n+\sum_{i=1,3,..,n-2} t_i u^i$$
where $t_i$ corresponds to giving vev to the tachyon in the i-th
twisted sector.  The fundamental field here is the chiral field
$Y$ where $u=e^{-Y}$. Let us set $t_i=0$.  The chiral ring in this case
is generated by $u$ modulo the equation $dW/dY=-udW/du=0$, which gives
the ring elements
$$u,u^2,...,u^{n-1}$$
in one to one correspondence with the twist fields of the $C/Z_n$
orbifold.  The fact that we are restricted to odd powers of $u$ is related
to the requirement of GSO projection which requires a $Z_2$ symmetry
under which $W\rightarrow -W$, which is realized here by $u\rightarrow -u$.
The charge of the field $u^k$
 is given by $k/n$, where we assign
charge
$1$ to $W$.  The charge in question here is the $U(1)$ charge
of the $N=2$ algebra.
More precisely the (left , right)-moving $U(1)$ charge of $u^k$ is given
by $(k/n,k/n)$.  In particular this state has total axial charge
$Q^5=F_L+F_R=2k/n$.
Note that the relevant ring of the LG model
when $u$ is the natural variable instead
of $Y$, is given by
$1,...,u^{n-2}$.  This would correspond to $N=2$ minimal model
 \ref\vw{C. Vafa
and N.P. Warner, ``Catastrophes and the Classification
of Conformal Theories,'' Phys. Lett. {\bf B 218} 51 (1989)}\ref\mart{
E. Martinec, ``Algebraic Geometry and
Effective Lagrangians,'' Phys. Lett. {\bf B 217} (1989) 431.}.

Aspects of chrial
rings in the conformal case have been studied
\ref\rings{W. Lerche, C. Vafa and N.P. Warner, ``Chiral Rings in $N=2$
Superconformal Theories,'' Nucl. Phys. {\bf B324}, 427 (1989). }. 
There is a 1-1 correspondence between chiral fields and ground states
of Ramond-Ramond sector:  For each chiral field of charge $Q$, there
is a Ramond ground state with left/right charge $q=Q-{\hat c}/2$, where $\hat c$
is the complex dimension of the conformal theory.  Note that
the spectrum in the Ramond ground state is symmetric under $q\rightarrow
-q$
as follows from CTP.
In the case of ${\bf C}/Z_n$ theory the (left or right) charges of the
Ramond
states are given by
$$({-1\over 2}+{1\over n},{-1\over 2}+{2\over n},...,{1\over 2}-{1\over n})$$
where we have used ${\hat c}=1$ for this case.  Note that the charges
of the ground states for $N=2$ minimal models also vary exactly
over this range, as follows from noting that ${\hat c}=1-{2\over n}$
in this case.  As discussed in detail in \rings\ for $N=2$
SCFT with discrete spectrum, the range of axial charges $Q^5$
of the Ramond ground state,
which is  left- plus right-moving
fermion number varies from $-{\hat c},...,+{\hat c}$.  In 
particular the absolute value of the highest axial charge in
the Ramond ground state is given by ${\hat c}$.  In the case of $C/Z_n$
the highest charge state is not ${\hat c}=1$, as that is not a compact
SCFT.  However its highest charge state is given by $1-{2\over n}$,
suggesting that this is the {\it effective} ${\hat c}$ for the tachyon
fields.

Now consider deforming the theory by turning on tachyon fields.
The worldsheet theory will flow in such a case.  For example suppose
we consider the deformation given by the superpotential
$$W=u^n+t u^k.$$
with $k<n$ and $k$ odd.  In this case the IR theory will be given
by $W=u^k$ which corresponds to the $C/Z_k$ theory.  It is natural
to ask how the chiral fields and the
corrsponding Ramond vacua of the $C/Z_n$ theory and those of $C/Z_k$
theory are related.  Note that by considering
isolated solutions of $udW/du=0$ we have $n-k$ massive vacua
which decouple at the IR. 
In addition we have
$k-1$ vacua which in the IR continue
to exist corresponding to the Ramond vacua of the $C/Z_k$ theory.
Note that the charges have now changed.
It was shown in \cecva\ how to follow the corresponding charges of the
vacua of the Ramond sector:  One studies the space of all the Ramond
vacua as a function of the deformation parameters.  This forms a 
holomorphic vector
bundle over the complex moduli space
$t_i$, which inherits a natural connection as considered
in the context of Quantum mechanics by Berry.  Let 
$$g_{i\overline j}=\langle i |{\overline j}\rangle$$
denote the overlap metric of the corresponding
Ramond ground states in a holomorphic
parameterization of the vacua (as defined in \cecva ).  Then the metric
satisfies a rich set of equations discovered in \cecva\ known
as $tt^*$ equations (topological /anti-topological equations).  The
main equation being that the curvature of this bundle is given by
$${\overline \partial_i}
(g\partial_j g^{-1})=-[{\overline C}_i,C_{ j}]$$
where $C_i$ denotes the matrix for the multiplication of the
Ramond ground states with the $i-$th chiral field.
Moreover, it was shown in \refs{\cecva ,\ansu}\ that the axial charge 
acting on the
Ramond ground states is related to the $tt^*$ metric by
$${1\over 2}[Q^5+m]|a\rangle =-(g\partial_{\tau} g^{-1})_a^b |b\rangle$$
where $Q^5$ denotes the axial charge ($Q^5=F_L+F_R$) and $m$ denotes
the total number of fields (which in the $C/Z_n$ case is just 1), and
$\tau$ denotes the RG paramater and corresponds to $W\rightarrow e^{\tau}W$. 
Note that $Q^5$ is not conserved during the flow and is conserved
only at the two ends of flow, where the theory becomes conformal. 
Moreover the $Q^5$ charges flow down in absolute value.  
It was shown in \ansu\ that $Q^5$ is also an ``index'' given by
$$Q^5_{ab}={i\tau\over L} {\rm Tr}_{ab} (-1)^F F e^{-\tau H}$$
where this is on a dual channel where space is of length $L$
(in the limit as $L\rightarrow \infty$) and $a,b$ denotes the left/right
vacua on the space.  $F$ here is the vectorial fermion number 
$F_L-F_R$ and
is related to the above axial charge by the change of channel
which converts axial charge to vector charge
(i.e. $Q^5 L\leftrightarrow iF\tau$).

\subsec{Proposal for the Potential Term}
We are now ready to present our proposal for the tachyon potential.
As noted in \cecva\
the axial charge matrix $Q^5$ is a natural matrix 
generalization of Zamolodchikov's c-function to the
case at hand.  In particular the absolute value of the highest eigenvalue
of $Q^5$ was proposed as a candidate
for a c-function defined for arbitrary massive $N=2$ theories.  Our proposal
for the tachyon potential as a function of $V(t_i,{\overline t}_i)$ is
\eqn\pro{V(t_i,{\overline t_i})={\rm max}|Q^5|.}
where the action involves $S\sim {1\over g_s^2} \int d^D x V$.
The most natural motivation for this conjecture is that the closed
string theory has a cosmological constant given by the central
charge of the theory, minus a constant.  In particular
$V\propto {\hat c}+{const.}$, is very natural.  The only
difference here is that the potential lives on a submanifold
of the spacetime, where the tachyon field is localized.  Thus
we have to find the effective ${\hat c}$ corresponding to the
tachyon degrees of freedom on the worldsheet.
 Since
${\rm max}|Q^5|$ effectively plays the role of 
the relevant piece of ${\hat c}$ for tachyonic
fields, therefore the above identification
is rather natural.  For the $C/Z_n$ vacuum the height of the tachyon
potential is given by
$$Q^5_{max}=(1-{2\over n})$$
This is in line with a recent proposal for the height of the tachyon
potential \atd.  The case studied there corresponds
to modding out the space by the rotation given by
$${\rm exp}(2\pi i J/N)$$
which to compare it to the case at hand we set $N=n/2$ to obtain
$$g={\rm exp}(4\pi i J/n)$$
It was argued that the tachyon potential in this case should be
$(1-{1\over N})$ which by the substitution $N\rightarrow n/2$ we get 
the above result.\foot{This
is presumably related to the fact that one has to take spin
connection rather than $SO(2)$ connection.}  In fact one can essentially map the logic
of the derivation of \atd\ to the above proposal:  The proposal
of \atd\ was based on identification of the tachyon potential
with the holonomy of the spin connection at infinity.  Consider
a string state in the twisted sector, and compute its vectorial fermion
number.  This is the same as the pull back of the holonomy of the spin
connection along this path.  In the limit the theory is the orbifold
we can take this path at any distance from the orbifold fixed point,
and in particular we can take it to correspond to the ground state of the
twisted sector.
Upon mirror symmetry, getting the $u$ field,
vectorial fermion
number gets converted to the axial
charge of the lowest
twisted state of the LG theory, which is the above proposal for the
tachyon potential.

Note that when we flow to the supersymmetric
theory, such as by considering 
$$W=u^n+t u$$
the $Q^5_{max}$ coming from the discrete states flow to zero,
as expected for the vanishing height of the potential
at the supersymmetric point.   In particular the height of the potential
is $1-{2\over n}$ only for $n>1$.

In our proposal for the tachyon potential \pro\ we have to search
for the maximum charge $Q^5$ among the states that can flow to zero
upon deformations.
Otherwise we would be also including supersymmetric states that have
zero potential.  

\subsec{Proposal for the Kinetic Term}

In case the target theory is supersymmetric, the fields $t_i$ correspond
to massless fields whose kinetic term is given as \cecva\
$$\int d^Dx G_{i\overline j}(t^k,{\overline t^k}) \partial_\mu
t^i \partial_\mu {\overline t^j}$$
where
$G_{i\overline j}$ is defined as follows.  Let $|\rho \rangle$
denote the ground state of the Ramond sector with lowest axial charge.
Suppose the superpotential involves $W=\int t_i \Phi_i$.  Define
$$|i\rangle =\Phi_i |\rho\rangle$$
Then
\eqn\defg{G_{i\overline j}={\langle {\overline j}|i\rangle \over \langle
{\overline \rho}|\rho\rangle}}
This definition of $G_{i\overline j}$
is simply the natural
normalized metric of the corresponding chiral fields.
In fact in the case the target is supersymmetric more is true:
$G$ is a Kahler metric, i.e. $G=\partial {\overline \partial}
K$ where $K={\rm log} \langle {\overline \rho}|\rho \rangle $
\ref\str{V. Periwal and A. Strominger, ``Kahler Geometry
of the Space of $N=2$ Superconformal Field Theories,''
Phys. Lett. {\bf B 235} (1990) 261.}\cecva .  

We propose the same kind of kinetic term also for tachyonic
deformations.  Namely again let $|\rho \rangle $ denote
the ground state of the Ramond sector with the lowest axial
charge.  By this we mean a normalizable such state.
For example for $C/Z_n$ orbifold, with mirror LG given by
$W=u^n$, the state $|\rho\rangle =|u\rangle$.  Of course
if we deform $W$, the state $|\rho \rangle $ may no longer
be given by $|u\rangle$ as the state with minimal $Q^5$ charge
may be a combination of various states.  However for deformations
of the form $W=u^n+tu$ it is easy to see by symmetry arguments
(as discussed in \cecva ) that $|u\rangle$ has the lowest axial charge
for all $t$.
 At any rate our proposal for the tachyon kinetic term
is 
$$K=\int G_{i\overline j}\partial_\mu t^i \partial_\mu {\overline t}^j,$$
with $G_{i\overline j}$ defined by \defg .
For example consider the case with $W=u^n+t u$.  In this case
the kinetic part of the action involving $t$ can be written as
\eqn\useful{\int 
{\langle {\overline u^2}|u^2\rangle\over
\langle {\overline u}|u\rangle } 
\partial_\mu t \partial_\mu {\overline t}}
To see this, note that since $t$ multiplies the field $u$
in the superpotential, and
since $|\rho\rangle =|u\rangle$ we have
$$u|\rho\rangle =u|u\rangle =|u^2\rangle$$
which yields from \defg\
$$G_{t{\overline t}}=
{\langle {\overline u^2}|u^2\rangle\over
\langle{\overline u}|u\rangle }. $$

In the non-supersymmetric case $G$ is no longer
a Kahler metric.
We are now ready to consider some examples.

\newsec{Examples}
In this section we consider some examples. Consider again ${\bf C}/Z_n$.
 The $tt^*$ geometry in this case is exactly the same
as the corresponding $N=2$ minimal model with a shift
$u^l\rightarrow u^{l+1}$, as discussed in the previous section.
For example, let us consider the ${\bf C}/Z_3$ case.  In this case
we have
$$W={1\over 3} u^3-t u$$
There are two ground states in the R sector with charges
$(-1/6,-1/6)$ and $(1/6,1/6)$ at $t=0$.  As $t\rightarrow \infty$
the corresponding charges, while opposite in sign, go to $0$.
   As shown in \cecva\ the
maximal charge $Q^5$, which we identify
with the Tachyon potential, can be computed using $tt^*$ equations
 and is given by
\eqn\defv{V=Q^5={-1\over 2} z {\partial v\over \partial z}}
where
$$z={4\over 3}|t|^{3/2}, \qquad |t|e^v={\langle 
{\overline u}^2|u^2\rangle
\over \langle {\overline u} |u\rangle }$$
and $v$ is a special solution to affine $A_1$ toda equation\foot{
This is a special case of Painleve III equation $Y''=(Y')^2/Y-Y'/z
+Y^3-1/Y$ where $Y^2=e^v$.}
\eqn\pth{v_{zz}+{v_z\over z}=4 {\rm sinh} (v)}
with  boundary conditions dictated by the regularity
of the solution and given by
$$v(z)\sim {-2\over 3}{\rm log} z+ s+ 2({3\over 4})^2
 e^s z^{4/3}+O(z^{8/3})\qquad z\rightarrow 0$$
$$v(z)\sim {1\over \sqrt{\pi z}}{\rm exp}(-2z) \qquad z>>0$$
where 
$$e^{s/2}={2^{2/3}}{\Gamma({2\over 3})\over \Gamma({1\over 3})}$$
Similarly the Kinetic term can be computed using \useful .  Putting
these together we obtain the tachyon action
$$S\sim {1\over g_s^2}\int |t|e^v\partial_\mu t \partial_\mu {\overline t}
+{-z\over 2}{\partial v\over \partial z}$$
(with $z={4\over 3}|t|^{3/2}$ and $v(z)$ defined above).

Let us see if this action has some of the expected properties:
Let us study the action near $t=0$.
At $t=0$ we have
$$\langle {\overline u} |u\rangle \big|_{t=0} =({3\over
4})^{-1/3}e^{-s/2}=1/
\langle {\overline u}^2|u^2\rangle \big|_{t=0}$$
Using this, if we identify the normalized tachyon field $T$
by $T=t \sqrt{{\langle {\overline u^2}|u^2\rangle\over
{\overline u}|u\rangle }}\big|_{t=0}$
we find that $S$ reduces near $T\sim 0$ to
$$S\sim {1\over g_s^2}\int \partial_\mu T \partial_\mu {\overline T}
+V(T,{\overline T})$$
where
$$V\sim {1\over 3}-{4\over 3} |T|^2+O(|T|^4) \qquad |T|\sim 0$$
Note that with this normalization the tachyon potential has the 
right height as well
as the correct negative mass squared for the tachyon
(we are using the convention $\alpha '=1$).  This is a rather
non-trivial
check of our proposal.

We would also like to study the solution for large $t$.  In this
limit we have
$$V\sim a |t|^{3/4}{\rm exp}(-{8\over 3}|t|^{3/2})\qquad |t|>>0.$$
where $a$ is a positive constant.  This potential approaches
0 exponentially fast for large $|t|$.

We would also like to prove
that $V$ has no extra critical points.  Note that $t=0 $ and
$t=\infty$ are two critical points and we do not expect
any other critical points.  In particular $dV/dt=0$ should have no
solution for finite $t>0$.   $V$ starts from a positive
value $1/3$ at $t=0$ and approaches $0$ at infinity, and we will
now show that it is monotonic.  If $dV/dt=0$ for some finite non-zero
$t$, it follows that $dV/dz=0$ for some $z>0$.
We know that
$${dV\over dz}={-1\over 2}{d\over dz}[{zd v\over d z}]
={-2z}\ {\rm sinh}(v)$$
where we used the definition of $V$ in \defv\ and the equation
satisfied by $v$ \pth .  For this to be zero, for non-zero $z$
we must have $v=0$ for some finite positive $z$.  $v>>0$ at $t=0$
and $v=0$ at $t=\infty$.  So if $v=0$ for some finite $z$ it must
become negative for finite values of $z$.  Given the fact that
$v$ approaches zero as $z\rightarrow \infty$ it implies
that there must be at least some critical points $v_z=0$ where $v<0$
and $v_{zz}>0$, which is incompatible with equation \pth .  This proves
that $V$ has no other critical points, as expected.

One can also ask how the tachyon field disappears in the $|t|>>0$
regime.  Note that in this limit
using $e^v\sim 1$ the kinetic term of the action becomes
$$\int |t|\partial_\mu t \partial_{\mu} {\overline t}$$
As $t$ rolls off to infinity, the above kinetic term has a
huge prefactor $|t|$, and this implies that for finite action
the field $t$ cannot vary appreciably over spacetime.  In other
words the field $t$ gets frozen out.   However this is not
very convincing, because by going to a field redefinition
namely $z$, which is the radial
part of the tachyon field,  we see that the kinetic term will become
$\partial z \partial z$ and the potential $a {\sqrt z}e^{-2z}$.
It would be interesting to study the meaning of the tachyon
field at the other end.

We can also consider other ${\bf C}/Z_n$ cases.  In all such
cases the $tt^*$ equations yields the tachyon
potential $V$. However an exact solution to $tt^*$
equations is not possible for general deformations.  For some
special cases of deformations the $tt^*$ equations are better
studied.  For example for
$$W=u^n+tu$$
the $tt^*$ equations boils down to affine $A_{n-2}$ toda equations
\cecva . This and some other special cases have been studied
in \refs{\cecva ,\ansu}\ to which we refer the interested reader.

We can also consider higher dimensional orbifolds
such as ${\bf C^2}\over Z_n$ studied in \refs{\posi ,\vata}.
Also in these case our proposal for the tachyon potential
$V$ has the right features to correspond to tachyon potential.
  For example for the non-supersymmetric
orbifold $({\rm exp}(2\pi ik_1/n),{\rm exp}(2\pi ik_2/n))$ we obtain
the height of the potential
$$V=2 (max_{l=1}^{n-1}{[lk_1]+[lk_2]\over n}-1)$$
where $[lk_i]=lk_i$ mod $n$ and $0\leq [lk_i]<n$.

For a generic
deformation all the charges go to zero and so $V\rightarrow 0$,
flowing to a supersymmetric vacuum.
Note that $V=0$ in the supersymmetric
case, which corresponds to $k_1=n-k_2$.  Apriori this
did not have to be the case, because our proposal for 
\pro\ is the maximum charge among all the states that can flow
to zero, and this is not the case for supersymmetric states.
Similarly one can consider the kinetic term and the variation of
potential term by studying solutions
to $tt^*$ equations.

\vskip 1cm

\centerline{\bf Acknowledgements}

We would like to thank K. Hori, A. Karch and A. Strominger for valuable
discussions.

This research is supported in part by NSF grants PHY-9802709
and DMS-0074329.

\listrefs

\end